\documentclass[conference,final]{IEEEtran}
\IEEEoverridecommandlockouts
\usepackage{cite}
\usepackage{amsmath,amssymb,amsfonts}
\usepackage{algorithm}
\usepackage{algorithmic}
\usepackage{graphicx}
\usepackage{textcomp}
\usepackage{multirow}
\def\BibTeX{{\rm B\kern-.05em{\sc i\kern-.025em b}\kern-.08em
    T\kern-.1667em\lower.7ex\hbox{E}\kern-.125emX}}

\begin{document}

\title{Unsupervised Time Series Extraction from Controller Area Network Payloads\\
\thanks{The views expressed in this document are those of the authors and do not reflect the official policy or position of the United States Air Force, the United States Army, the United States Department of Defense or the United States Government. Approved for public release (Case Number 88ABW-2018-1916).}
}

\author{\IEEEauthorblockN{Brent J. Stone}
	\IEEEauthorblockA{\textit{Department of Electrical and Computer Engineering} \\
		\textit{Air Force Institute of Technology}\\
		WPAFB, OH USA \\
		brent.jk.stone@gmail.com}
	\and
	
	\IEEEauthorblockN{Scott Graham}
	\IEEEauthorblockA{\textit{Department of Electrical and Computer Engineering} \\
		\textit{Air Force Institute of Technology}\\
		WPAFB, OH USA \\
		scott.graham@afit.edu}
	\and
	
	\IEEEauthorblockN{Barry Mullins}
	\IEEEauthorblockA{\textit{Department of Electrical and Computer Engineering} \\
		\textit{Air Force Institute of Technology}\\
		WPAFB, OH USA \\
		barry.mullins@afit.edu}
	\and
	
	\IEEEauthorblockN{Christine Schubert Kabban}
	\IEEEauthorblockA{\textit{Department of Mathematics and Statistics} \\
		\textit{Air Force Institute of Technology}\\
		WPAFB, OH USA \\
		christine.schubert@afit.edu}
}

\maketitle

\begin{abstract}
This paper introduces a method for unsupervised \textit{tokenization} of Controller Area Network (CAN) data payloads using bit level transition analysis and a greedy grouping strategy. The primary goal of this proposal is to extract individual time series which have been concatenated together before transmission onto a vehicle's CAN bus. This process is necessary because the documentation for how to properly extract data from a network may not always be available; passenger vehicle CAN configurations are protected as trade secrets. At least one major manufacturer has also been found to deliberately misconfigure their documented extraction methods. Thus, this proposal serves as a critical enabler for robust third-party security auditing and intrusion detection systems which do not rely on manufacturers sharing confidential information.
\end{abstract}


\section{Introduction} \label{intro}
Current production vehicles are becoming as much software as they are hardware. Their networks now feature optional persistent Internet connections and are complex enough to support emerging technologies such as autonomous driving and Vehicle-to-Everything (V2X) applications \cite{b1}. Mass production of vehicles with Internet accessible computers capable of controlling all aspects of the vehicle makes incorporating and validating \textit{defense in depth} cyber security techniques a practical necessity. 

Bug bounty programs, Cyber Emergency Response Teams (CERT), and widely attended 'hacker' conferences are all strong evidence that independent research is an essential part of developing and validating robust cyber security practices. We assume that the computing systems and networks used in the automotive industry are no exception to needing third party security auditing to establish and improve robust \textit{defense in depth} security measures. Third party research also ensures that accidental or deliberate network flaws such as the 2015 Volkswagen emission scandal are identified and corrected more quickly \cite{vw}. To that end, we intend the methods presented in this paper to address the absence of a CAN payload \textit{tokenization} technique. Without an effective payload \textit{tokenization} technique, third party research is limited to manual reverse engineering a small set of vehicles and hoping those are representative of the broader market, using methods which ignore the useful information present in CAN payloads, or somehow gain access to confidential manufacturer specifications.

\section{Background} \label{background}

\subsection{Automated Network Traffic Reverse Engineering}

The concept of automated protocol reverse engineering using observed network traffic is an active area of research. However, practically all published research is focused on analyzing a heterogeneous mix of text-based application layer protocols with the goal of facilitating \textit{deep packet inspection} \cite{b5, b11, b14, b15, b17}. The approach presented in this paper is based on analyzing payloads of a single known protocol-Controller Area Network (CAN)-which is not text-based. The key difference is the difficulty of \textit{lexical analysis}.

We propose the difference between translating sentences written using Japanese Katakana and English is a reasonable analogy to the difference between existing research and the problem addressed by this paper. Automated translation of either language certainly shares similarities once the words in a sentence and their ordering have been identified. However, with text-based network protocols and English there is a finite set of delimiters that are almost always present between `words'. Thus, the \textit{lexical analysis} phase proposed in \cite{b5, b11, b14, b15, b17} and related work is almost always a trivial process using a set of delimiter characters known \textit{a priori}. Sentences written with Japanese Katakana and CAN payloads do not use explicit delimiters. This makes `word' discovery non-trivial in these contexts.

The approach proposed by Markovitz and Wool is the only published method found to address the problem of automated reverse engineering of CAN protocol payloads \cite{b20}. Markovitz and Wool proposed a brute force search followed by heuristic selection using the number of unique values present in each \textit{time series} considered. The authors reported that this lexical analysis method had poor accuracy using self generated network data. Thus, the reverse engineering pipeline presented in this paper is assumed to be the first proposal for robust automated reverse engineering of non-text network protocol payloads.

\subsection{Tokenization}

The term \textit{tokenization} is taken from compiler design in computer science. Compilers are the software which converts a program into a series of operations that can run on computer hardware \cite{b18}. \textit{Lexical analysis} is the first step of a compiler which uses human readable programming code as an input. The \textit{tokenization} process identifies the individual logical units, or \textit{tokens}, that code consists of. For example, the following program code results in the nine \textit{tokens} \textit{`for', `x', `in', `range', `(', `0', `,', `10', `)'}:
\begin{align*}
for \; x \; in \; range(0, 10)
\end{align*}

If the \textit{f} and \textit{o} in the token \textit{`for'} are incorrectly separated during \textit{tokenization}, then the following steps in the compiler will fail. The compilation should also fail if the tokens \textit{`for'} and \textit{`x'} are \textit{not} separated during \textit{tokenization}.

We define the \textit{tokenization} of CAN data as the process of identifying the logically distinct \textit{time series} present within message payloads using the same arbitration ID. The term \textit{time series} is taken from the National Institute of Science and Technology (NIST) definition of a univariate sequence of values ordered by the time observed \cite{b19}. Examples of \textit{time series} in a vehicle might be measurements by an Electronic Control Unit (ECU) monitoring the front right wheel's rotations per minute (RPM), steering wheel angle, or engine RPM. We will refer to individual \textit{time series} extracted through \textit{tokenization} simply as \textit{signals} for the remainder of this paper.

\begin{figure}[!t]
	\centering
	\includegraphics[width=3.4in]{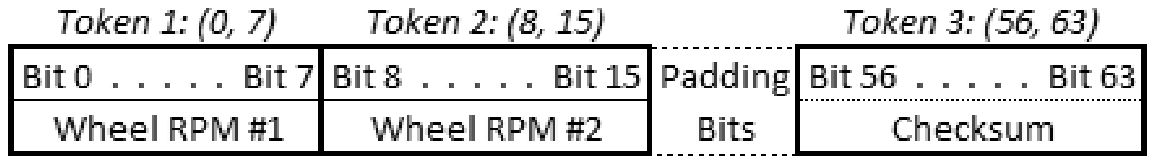}
	\caption{Example of a CAN payload tokenization}
	\label{tokens}
\end{figure}

As an example, imagine the RPM \textit{signal} for two of a vehicle's wheels and a checksum are all contained in the set of 64-bit payloads using a CAN arbitration ID of 0xA15. The two RPM measurements and checksum are 8-bit \textit{signals}. A possible \textit{tokenization} would be the set of start and stop indices: (0, 7), (8, 15), (56, 63). The \textit{bit positions} 16 through 55 are \textit{padding bits} which are consistently 1 or 0 in every observed payload using ID 0xA15. Figure \ref{tokens} depicts this hypothetical \textit{tokenization} scenario.

We empirically found that \textit{tokenization} is necessary to correctly interpret CAN message payloads. This is because a series of payloads using a shared arbitration ID often contains multiple sensor readings concatenated together. This observation is echoed by other third party CAN research findings \cite{b20, b21}.  Thus, we define the input of CAN \textit{tokenization} as a series of chronologically-ordered CAN message payloads present in a sample of CAN network traffic which share the same arbitration ID. We assume payloads for each ID always use the same bit-width (e.g., the payload is always 64-bits) and logical formatting. This assumption is based on our empirical analysis of eight vehicles produced for the United Sates market and the findings of Miller and Valasek \cite{b21}. The output of CAN \textit{tokenization} is the set of \textit{bit positions} within the payload that bound each logically-distinct \textit{signal}. 

Correct payload \textit{tokenization} and classification of CAN \textit{signals} enables a broad range of findings. For example, extracting the brake pedal position \textit{signal} from a CAN bus is sufficient to identify who is driving the vehicle out of a population of known drivers \cite{fingerprint}. Using the signal type, transmission frequency, and other features may be sufficient to fingerprint specific Electronic Control Unit (ECU) hardware in a similar fashion. Automating the process of fingerprinting particular ECUs could lead to rapidly, passively, and cheaply identifying vehicles affected by published ECU firmware vulnerabilities. Again, the first step to achieving such results is the \textit{tokenization} of CAN payloads.

\section{Research Method}\label{method}

\subsection{Transition Analysis}\label{aggregation}
The goal of this initial CAN payload \textit{tokenization} proposal is to correctly extract continuous numerical \textit{signals} transmitted over a vehicle's CAN bus. We assume the preponderance of payloads produced in production CAN networks are mostly comprised of continuous and categorical data. This assumption is again based on empirical research of eight production vehicles and the work done by Miller and Valasek \cite{b21}. Extracting continuous numerical \textit{signals} from a heterogeneous population of continuous and categorical \textit{signals} achieves three important objectives. First, it provides the continuous numerical \textit{signals} as a ready-to-analyze output. Second, removing these \textit{signals} reduces the bit width of the remaining payload segments which need to be \textit{tokenized}. Third, removing continuous data from observed CAN payloads allows methods targeted for the \textit{tokenization} of categorical data to operate with the assumption that the data set is a homogeneous population of categorical data.

The reason continuous numerical \textit{signals} can and should be targeted first is because there's a predictable relationship between \textit{bit positions} used to convey continuous numerical data. Numerical data can be represented with a binary protocol like CAN using a range of encoding schemes such as \textit{unsigned} values or signed values using \textit{two's compliment, one's compliment, signed magnitude}, and more. The common feature of these various encoding schemes is the notion that bits are ordered from a \textit{least significant bit} (LSB) to a \textit{most significant bit} (MSB). The LSB represents the $2^0$'s place and the MSB represents the $2^{n-1}$'s place where \textit{n} is the bit width being used.

We empirically found that vehicle sensors sampling continuous real world processes such as velocity, pedal position, and steering angle many times a second using numerical data will produce approximately continuous numerical \textit{time series}. To say this another way, vehicle sensors measuring locomotion will report numbers that have small differences between sequential samples. RPM will not jump between 1,200, 7,000, back down to 2,000, and then 5,000 within one second unless the engine might be exploding. Rather, a generally smooth increase from one value to another will be observed such as 2,000 to 2,032 and then 2,053 RPM.

The use of bit ordering from LSB to MSB and the approximately continuous numerical nature of \textit{signals} produced by locomotion related ECUs causes predictable relationships to form between neighboring \textit{bit positions} within CAN payloads. Transition analysis can quantify this predictability for unsupervised payload \textit{tokenization}.

A \textit{bit position} transitions when it flips between 1 and 0 in chronologically-sequenced CAN payloads using the same arbitration ID and bit width. Bit level transition analysis can be efficiently calculated by storing observed payloads into an $M \times N$ boolean matrix. $M$ is the number of row vectors with one row per observed CAN message payload. $N$ is the bit width of the payloads with column vectors representing the relative \textit{bit positions} within the payloads. See Table \ref{table-boolean} for an example of a 10 x 8 boolean matrix representing 10 samples of an 8-bit payload.

By performing an exclusive or (XOR) of each sequential pair of row vectors in such a boolean matrix, a \textit{transition matrix} is the created with $M-1$ rows and 1s anywhere a bit transition occurred. Table \ref{table-transitions} is the \textit{transition matrix} produced from Table \ref{table-boolean}. In this example the \textit{0th} row vector is XORed with the \textit{1st} row vector.

\begin{table}[!t]
	\centering
	\caption{Example Boolean Matrix of Payloads for a Single Arb ID}
	\label{table-boolean}
	\begin{tabular}{|r|l|l|l|l|l|l|l|l|}
	\hline
	\multicolumn{1}{|c|}{\multirow{2}{*}{Observation}} 		 & 
	\multicolumn{8}{c|}{\textbf{Bit Position}}                                                             \\ \cline{2-9} 
	\multicolumn{1}{|c|}{} & $\boldsymbol{2^7}$ & $\boldsymbol{2^6}$ & $\boldsymbol{2^5}$ & $\boldsymbol{2^4}$ & $\boldsymbol{2^3}$ & $\boldsymbol{2^2}$ & $\boldsymbol{2^1}$ & $\boldsymbol{2^0}$  \\ \hline
	\textit{\textbf{0}} & 0          & 0          & 0          & 0          & 0          & 0          & 0          & 0           \\ \hline
	\textit{\textbf{1}} & 0          & 0          & 0          & 0          & 0          & 0          & 0          & 1           \\ \hline
	\textit{\textbf{2}} & 0          & 0          & 0          & 0          & 0          & 0          & 1          & 0           \\ \hline
	\textit{\textbf{3}} & 0          & 0          & 0          & 0          & 0          & 0          & 1          & 1           \\ \hline
	\textit{\textbf{4}} & 0          & 0          & 0          & 0          & 0          & 1          & 0          & 0           \\ \hline
	\textit{\textbf{5}} & 0          & 0          & 0          & 0          & 0          & 1          & 0          & 1           \\ \hline
	\textit{\textbf{6}} & 0          & 0          & 0          & 0          & 0          & 1          & 1          & 0           \\ \hline
	\textit{\textbf{7}} & 0          & 0          & 0          & 0          & 0          & 1          & 1          & 1           \\ \hline
	\textit{\textbf{8}} & 0          & 0          & 0          & 0          & 1          & 0          & 0          & 0           \\ \hline
	\textit{\textbf{9}} & 0          & 0          & 0          & 0          & 1          & 0          & 0          & 1           \\ \hline 
\end{tabular}
\end{table}

\begin{center}
	\medskip
	\begin{tabular}{ccccccccc}
				 & 0 & 0 & 0 & 0 & 0 & 0 & 0 & 0 \\
		$\oplus$ & 0 & 0 & 0 & 0 & 0 & 0 & 0 & 1 \\
				 \hline
				 & 0 & 0 & 0 & 0 & 0 & 0 & 0 & 1 \\
	\end{tabular}
	\medskip
\end{center}

The \textit{1st} row vector is XORed with the \textit{2nd} row vector and so on for all sequential row vectors in the boolean matrix. Summing the 1s in each column vector (\textit{bit position}) of the \textit{transition matrix} produces a $1 \times N$ row vector. For the remainder of this proposal this summary row vector will be referred to as a \textit{Transition Aggregation N-Gram} (TANG).

\begin{table}[!t]
	\centering
	\caption{Example Transition Matrix and Transition Aggregation}
	\label{table-transitions}
	\begin{tabular}{|r|l|l|l|l|l|l|l|l|}
	\hline
	\multicolumn{1}{|c|}{\multirow{2}{*}{XOR Result}} & \multicolumn{8}{c|}{\textbf{Bit Position}}                                                             \\ \cline{2-9} 
	\multicolumn{1}{|c|}{}                                   & $\boldsymbol{2^7}$ & $\boldsymbol{2^6}$ & $\boldsymbol{2^5}$ & $\boldsymbol{2^4}$ & $\boldsymbol{2^3}$ & $\boldsymbol{2^2}$ & $\boldsymbol{2^1}$ & $\boldsymbol{2^0}$  \\ \hline
	\textit{\textbf{Obs. $0 \oplus 1$}}  & 0          & 0          & 0          & 0          & 0          & 0          & 0          & 1           \\ \hline
	\textit{\textbf{Obs. $1 \oplus 2$}}  & 0          & 0          & 0          & 0          & 0          & 0          & 1          & 1           \\ \hline
	\textit{\textbf{Obs. $2 \oplus 3$}}  & 0          & 0          & 0          & 0          & 0          & 0          & 0          & 1           \\ \hline
	\textit{\textbf{Obs. $3 \oplus 4$}}  & 0          & 0          & 0          & 0          & 0          & 1          & 1          & 1           \\ \hline
	\textit{\textbf{Obs. $4 \oplus 5$}}  & 0          & 0          & 0          & 0          & 0          & 0          & 0          & 1           \\ \hline
	\textit{\textbf{Obs. $5 \oplus 6$}}  & 0          & 0          & 0          & 0          & 0          & 0          & 1          & 1           \\ \hline
	\textit{\textbf{Obs. $6 \oplus 7$}}  & 0          & 0          & 0          & 0          & 0          & 0          & 0          & 1           \\ \hline
	\textit{\textbf{Obs. $7 \oplus 8$}}  & 0          & 0          & 0          & 0          & 1          & 1          & 1          & 1           \\ \hline
	\textit{\textbf{Obs. $8 \oplus 9$}}  & 0          & 0          & 0          & 0          & 0          & 0          & 0          & 1           \\ \hline \hline 
	\multicolumn{1}{|l|}{\textit{\textbf{TANG}}}      & \textbf{0} & \textbf{0} & \textbf{0} & \textbf{0} & \textbf{1} & \textbf{2} & \textbf{4} & \textbf{9} \\ \hline
\end{tabular}
\end{table}

\subsection{Greedy Bit-Position Grouping} \label{grouping}

\textit{Bit positions} with the largest transition count in a TANG might be the LSB of a numerical \textit{signal} within the CAN payloads. If a LSB and its neighboring \textit{bit positions} represent a monotonically-decreasing gradient of transition counts in a TANG, this is evidence that they belong to the same continuous numerical \textit{signal}. This behavior is demonstrated in the TANG produced from Table \ref{table-boolean}. \textit{Bit position} $2^0$ was the LSB of the \textit{bit positions} representing the unsigned integer sequence counting from 0 in row 0 to 9 in row 9.

Algorithm \ref{alg-tokenize} presents a greedy strategy for clustering \textit{bit positions} suspected of being a continuous numerical \textit{signal} using TANGs. The benefits of this greedy approach are the ability to work with the univariate format of TANGs, no requirement to specify the number of \textit{signals} in a payload, and no reliance on heuristics or \textit{a priori} knowledge of the CAN payload. It is possible to implement Algorithm \ref{alg-tokenize} without sorting a copy of the TANG or using nested loops; however, this slightly more inefficient version is presented to allow for a conceptually straightforward written explanation. 

Algorithm \ref{alg-tokenize} begins by sorting a TANG by observed transition count on line 6. This sorted list of \textit{bit positions} is placed in a stack (a \textit{last in-first out} data structure) with \textit{bit positions} that transitioned the most frequently at the top of the stack. The stack is iteratively \textit{popped} on lines 9 and 10 until all \textit{bit positions} have been considered. When a \textit{bit position} is \textit{popped} from the top of the stack, the conditional statement on line 11 uses the \textit{`complete'} list to check if it is already clustered. If not, the assumption is made that this \textit{bit position} is the \textit{least significant bit} (LSB) of a \textit{signal}. Lines 12 through 15 create a new \textit{cluster} with this \textit{bit position}. 

The nested loop on lines 16 through 21 then consider all \textit{bit positions} on the left-hand or right-hand side (\textit{endian} dependent) of the new LSB. These neighbor \textit{bit positions} are added to the new \textit{cluster} of \textit{bit positions} while they represent a monotonically-decreasing gradient of transition values in the TANG. The `\textit{less than or equal}' transition count comparison on line \ref{tunable} could be replaced with an adjustable maximum difference threshold. Once all \textit{bit positions} have been considered, the clusters are returned as output. Each cluster of \textit{bit positions} represent an educated guess about where continuous numerical \textit{signals} exist in the arbitration ID's payloads. The maximum difference threshold method was used when producing findings and examples for this paper. 

\setlength{\textfloatsep}{6pt}
\begin{algorithm}[!t]
	\caption{Greedy Payload \textit{Tokenization} Using Its TANG}\label{alg-tokenize}
	\begin{algorithmic}[1]
		\REQUIRE A 2xN matrix with one row of bit position indices and a corresponding row for the TANG. The columns represent the \textit{0th} to \textit{n-1} bit positions of a particular arbitration ID's CAN payloads.
		
		\IF{data payload is assumed to use big-endian}
		\STATE{offset $\leftarrow -1$}
		\ELSE 
		\STATE{offset $ \leftarrow 1$}
		\ENDIF
		
		\STATE stack $\leftarrow sort\_by\_transition\_count(TANG)$
		\STATE clusters $\leftarrow [\;]$
		\STATE complete $\leftarrow [\;]$
		
		\WHILE{stack \NOT empty} \label{outer}
		\STATE current $\leftarrow$ stack.pop
		\IF{current.index \NOT \textbf{in} complete}
		\STATE cluster $\leftarrow [\;]$
		\STATE cluster.append(current)
		\STATE complete.append(current.index)
		\STATE neighbor $\leftarrow$ TANG[current.index + offset]
		\WHILE{neighbor.transitions $\leq$ current.transitions} \label{tunable} 
		\STATE cluster.append(neighbor)
		\STATE complete.append(neighbor.index)
		\STATE current $\leftarrow$ neighbor
		\STATE neighbor $\leftarrow$ TANG[neighbor.index + offset]
		\ENDWHILE
		\STATE clusters.append(cluster)
		\ENDIF
		\ENDWHILE
		\RETURN clusters	
	\end{algorithmic}
\end{algorithm}

\section{Findings} \label{findings}

\subsection{Anecdotal Results of Greedy CAN Payload Tokenization} \label{anecdotes}
In this section several anecdotal examples of TANGs and the results of Algorithm \ref{alg-tokenize} are presented based upon approximately 10 minutes of CAN network traffic collected from a 2012 model year minivan being operated in \textit{city driving} conditions. This vehicle is one of eight model year 2008 or later passenger vehicles studied. This sample population of vehicles includes sedans, sport utility vehicles, pickup trucks, and minivans. These vehicles used traditional gasoline internal combustion or hybrid powertrains; no diesel vehicles were studied. Two vehicles were equipped with a manual transmission. While the CAN network in each vehicle studied is at least superficially unique, we found Algorithm \ref{alg-tokenize} achieved similar success across all of the vehicles. Due to space limitations, only anecdotal results from one vehicle will be presented.

These findings are deliberately presented as anecdotal results as opposed to a qualitative evaluation using synthesized data. Providing specific qualitative performance statistics for Algorithm \ref{alg-tokenize} using synthetic (but known) CAN traffic is unhelpful at best and misleading at worst. The fundamental problem being addressed by this paper is a lack of \textit{a priori} knowledge of the CAN network beyond the CAN protocol specification. Creating a testbed CAN network and explicitly or implicitly claiming it is representative of all production vehicle CAN networks for the purposes of validating Algorithm \ref{alg-tokenize} is a non-trivial claim. Unfortunately, further exploring the interesting problem of creating a sufficiently `representative' CAN network is also beyond the scope and length limits of this paper.

%\begin{figure}[!t]
%	\centering
%	\includegraphics[width=3.4in]{easy_1.png}
%	\caption{An Example of an Easily Tokenized Payload}
%	\label{easy1}
%\end{figure}

\begin{figure}[!t]
	\centering
	\includegraphics[width=3.4in]{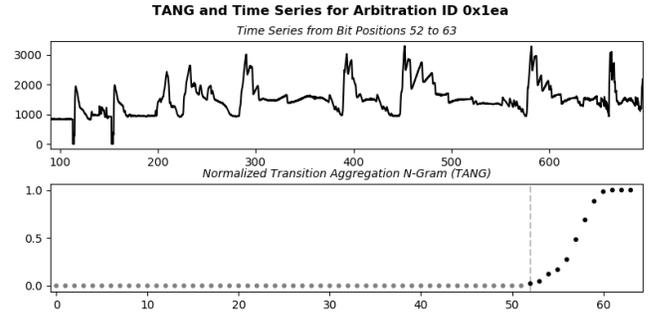}
	\caption{An Example of an Easily Tokenized Payload}
	\label{easy2}
\end{figure}

\begin{figure}[!t]
	\centering
	\includegraphics[width=3.4in]{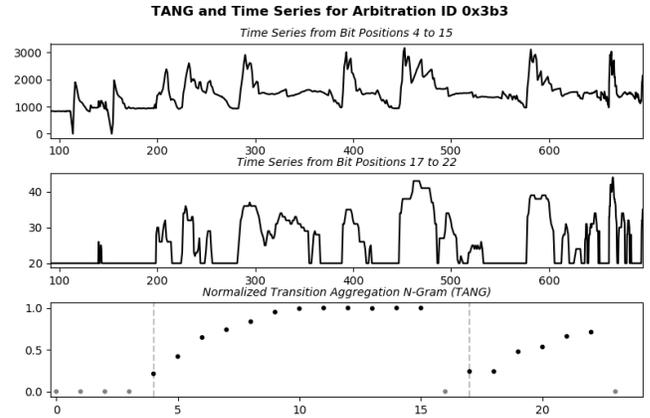}
	\caption{A Payload With Two Time Series Concatenated Together}
	\label{easy3}
\end{figure}

%\begin{figure}[!t]
%	\centering
%	\includegraphics[width=3.4in]{easy_4.png}
%	\caption{A Payload With Three Signals. The Signal at Bits 17 to 23 Closely Resemble Fig.~\ref{easy3}'s Signal at Bits 17 to 22}
%	\label{easy4}
%\end{figure}

Figures \ref{easy2} and \ref{easy3} are examples of CAN payloads with continuous numerical \textit{signals} targeted by Algorithm \ref{alg-tokenize}. With the exception of the bottom plot in each figure, these plots represent each logically distinct \textit{time series} present in the payloads of the listed arbitration IDs. These \textit{time series} plots represent a non-overlapping subset of \textit{bit positions} present in the total payload size shown in the TANG plot at the bottom of each figure. The vertical axis of these \textit{time series} plots is the unsigned integer interpretation for the indicated cluster of \textit{bit positions} within each payload. The horizontal axis is the chronological index of the payloads observed in the sample. Thus, these \textit{time series} plots can be read from left to right as the unsigned integer value that cluster of \textit{bit positions} took on as time progressed in the driving sample.

The TANG plot at the bottom of each figure is a graphical representation of the TANG for the listed arbitration ID. The vertical axis of this TANG plot is the min-max normalized transition count (transitions divided by total observations) for each \textit{bit position} in the eight byte payloads. Higher values on this vertical axis indicate the \textit{bit position} marked by the horizontal axis transitioned more frequently. The horizontal axis indicates the total bit positions in the series of observed payloads.

Vertical dashed lines indicate the \textit{most significant bit} (MSB) of a \textit{signal} identified by Algorithm \ref{alg-tokenize}. The LSB of each \textit{signal} is not explicitly identified to avoid clutter. However, both the LSB and MSB are explicitly listed in the sub-title of each \textit{time series} plot. Grey points in the TANG plot indicate possible \textit{padding bits} observed in the CAN data sample; these \textit{bit positions} never transitioned in the driving sample.

\begin{figure}[!t]
	\centering
	\includegraphics[width=3.4in]{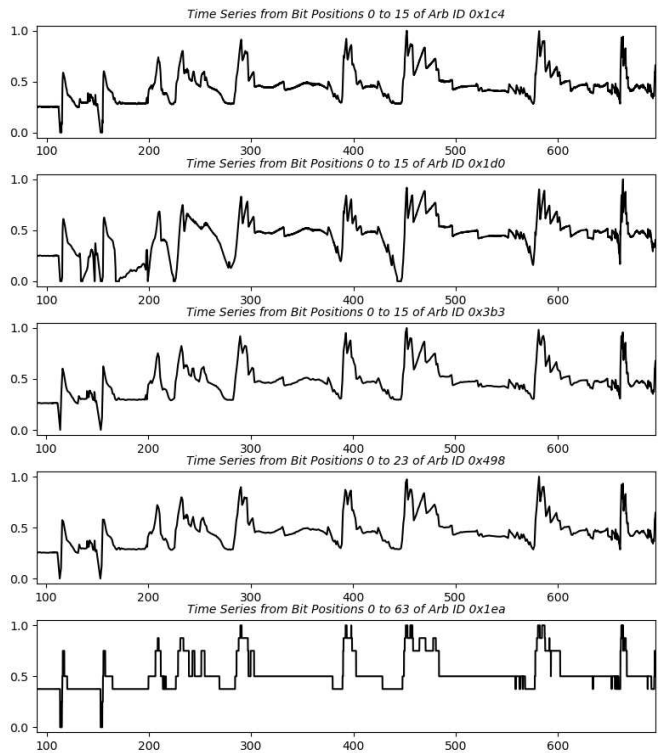}
	\caption{An Example of Tokenization Consistency and Time Series Similarities}
	\label{cluster}
\end{figure}

Figure \ref{cluster} is an example of how \textit{time series} similar to those seen \figurename{~\ref{easy2}} and \figurename{~\ref{easy3}} are consistently \textit{tokenized} by Algorithm \ref{alg-tokenize} across multiple arbitration IDs in a production vehicle. We found the phenomenon of similar \textit{time series} being present to occur in all eight vehicles studied.

\section{Future Work} \label{future}
In the future we will present an unsupervised pipeline for identifying and clustering continuous numerical \textit{signals} expected to be correctly \textit{tokenized} by the proposed \textit{tokenization} strategy. This pipeline was used to generate the \textit{signal} cluster in \figurename{~\ref{cluster}}. The pipeline will be used to rapidly produce a large data set of accurately \textit{tokenized time series} present in production CAN networks. That empirical data set will be used to formulate a `gold standard' labeled data set as part of a proposal for robust validation of \textit{tokenization} or intrusion detection algorithms for cyber-physical systems using CAN.

\section{Conclusion}
This paper introduced the idea of CAN payload \textit{tokenization} and motivated the need for such a pipeline. Section \ref{method} proposed an efficient method of quantifying predictable bit level relationships in CAN payloads using \textit{Transition Aggregation N-Grams} (TANGs). A greedy strategy was proposed as a proof of concept for how TANGs can be used to automate CAN payload \textit{tokenization}. Section \ref{findings} presents three examples of Algorithm \ref{alg-tokenize}'s performance with real world CAN data.

Payload \textit{tokenization} techniques are sorely needed for third-party research in domains using CAN and similar protocols. This proposal partially addresses that shortfall.

\end{document}